\documentclass[page-classic]{epl2} 

\usepackage{graphicx}
\usepackage{lipsum}
\usepackage{float}
\usepackage{amsmath}
\usepackage{amssymb}

\title{
Analytical results for the distribution of 
shortest path lengths in random networks
}
\shorttitle{The distribution of shortest path lengths in random networks} 

\author{Eytan Katzav\inst{1} \and Mor Nitzan\inst{1,2} \and Daniel ben-Avraham\inst{3} 
\and P.L. Krapivsky\inst{4} \and Reimer K\"uhn\inst{5} \and Nathan Ross\inst{6} \and Ofer Biham\inst{1}}
\shortauthor{Eytan Katzav \etal}

\institute{                    
  \inst{1} Racah Institute of Physics, The Hebrew University, Jerusalem 91904, Israel\\
  \inst{2} Department of Microbiology and Molecular Genetics, Faculty of Medicine, 
The Hebrew University, Jerusalem 91120, Israel\\
  \inst{3} Department of Physics, Clarkson University Potsdam, NY 13699-5820, USA\\
  \inst{4} Department of Physics, Boston University, Boston, MA 02215, USA\\
  \inst{5} Department of Mathematics King's College London Strand, London WC2R 2LS, UK\\
  \inst{6} School of Mathematics and Statistics, University of Melbourne, VIC 3010, Australia
}

\abstract{
We present two complementary analytical approaches for calculating the 
distribution of shortest path lengths in Erd{\H o}s-R\'enyi networks,
based on recursion equations for the shells around a reference
node and for the paths originating from it. 
The results are in agreement with numerical
simulations for a broad range of network sizes and connectivities.
The average and standard deviation of the distribution are also obtained.
In the case that
the mean degree scales as 
$N^{\alpha}$ 
with the network size,
the distribution becomes extremely narrow
in the asymptotic limit, namely almost all
pairs of nodes are equidistant, at distance 
$d=\lfloor 1/\alpha \rfloor$
from each other.
The distribution of shortest path lengths between nodes of degree $m$
and the rest of the network is calculated.
Its average is shown to be a monotonically decreasing function of $m$,
providing an interesting relation between a local property and a global
property of the network.  
The methodology presented here can be applied to more general classes
of networks.
}

\begin{document}

\maketitle




The increasing interest in network research in recent years is motivated by 
the realization that a large variety of systems and processes
which involve interacting objects can be described
by network models
\cite{Barabasi2002,Caldarelli2007,Newman2010,Estrada2011}.
In these models, the objects are represented by nodes and
the interactions are expressed by edges.
Pairs of connected nodes can affect each other directly.
However, the interactions between most pairs of nodes 
are indirect, mediated by intermediate nodes and edges.
Important properties of these indirect interactions such as their 
strengths, delay times, coordination, correlation and synchronization 
depend on the paths between different nodes.
A pair of nodes, $i$ and $j$, may be connected by a large number
of paths. The shortest among these paths are of
particular importance because they are likely to provide the fastest 
and strongest interaction between these two nodes.
Therefore, it is of interest to study the distribution of shortest 
path lengths (DSPL) between nodes in different types of networks.
Such distributions are expected to depend on the network 
structure and size.

Random networks of the Erd{\H o}s-R\'enyi (ER) 
type were studied extensively
since the 1950's
\cite{Erdos1959,Erdos1960,Erdos1961} 
using mathematical methods and computer simulations
\cite{Bollobas2001}.
The increasing availability of empirical data on networks since
the late 1990's stimulated much theoretical interest, 
leading to new results
for ER networks 
\cite{Hofstad2005,Debacco2015}.
Measures such as the diameter and the average path length
were studied extensively
\cite{Watts98,Fronczak2004}.
However, apart from a few studies,
the entire DSPL
has attracted 
little attention
\cite{Blondel2007,Dorogotsev2003,Esker2008}.
This distribution 
is of great importance for the 
temporal evolution of dynamical processes on networks, 
such as signal propagation, navigation and epidemic spreading 
\cite{Satorras2001}.
It determines the number of nodes
exposed to a propagating signal originated 
from a given node as a function of time.
More generally, the
shortest paths can be considered as the backbone of a more complete
set of paths between pairs of nodes. While the shortest paths provide
the fastest propagation, signals also utilize longer paths which are
more numerous.
This was demonstrated in studies of first passage times in diffusive
processes on networks
\cite{Sood2005}.  

In this Letter we present two analytical approaches for calculating 
the 
DSPL
between nodes in the ER network,
referred to as the recursive shells approach (RSA)
and the recursive paths approach (RPA).
Using recursion equations we study 
this distribution
in different regimes, namely sparse and dense networks
of small as well as asymptotically large sizes.
Consider an ER network of $N$ nodes, where each pair of nodes
is independently connected with probability $p$. 
We denote such a network by ER($N,p$).
Its degree sequence follows the
Poisson distribution with the parameter 
$Np$, 
which is equal to the average degree.
Such networks are often studied in the asymptotic limit, where
$N \rightarrow \infty$.
In this limit, one can identify different regimes,
according to the scaling of $p$ vs. $N$.

For sparse networks, denoted by 
ER($N,c/N$),
the average degree is
$c=Np$. 
At $c=1$ there is a percolation transition.
For $c<1$, the network consists of small
isolated clusters. 
For $c>1$, a giant component 
of size which scales linearly with $N$ is formed,
in addition to the small, isolated components of maximal 
size which scales as $\ln N$
\cite{Bollobas2001}.
For dense networks, the parameter
$p$ scales as $N^{\alpha-1}$,
where 
$0 < \alpha < 1$,
the mean connectivity grows with the network size
as
$N^{\alpha}$
and the number of isolated components vanishes.

When a pair of nodes resides on the same connected sub-network, one can
identify paths connecting these nodes. 
The path length is the number of edges along
the path.
The distance 
$d_{ij}$
between a pair of different nodes $i$ and $j$
is the length of the shortest path connecting them.
When $i$ and $j$ reside on different sub-networks,
there is no path between them
and thus
$d_{ij} \equiv \infty$.
The tail-distribution
$F_N(k) = Pr(d>k)$, $k=0,1,2,\dots,N-1$,
is the probability that the distance $d$ between a random pair
of nodes 
in an ER network of size $N$
is larger than $k$.
Clearly, the probability that two distinct random 
nodes are at a distance $d>0$ from each other is 
$F_N(0)=1$,
while the probability that
$d>1$,
namely they are not directly connected, 
is
$F_N(1)=q$,
where $q=1-p$.
The probability distribution 
$P_N(k)$ 
can be recovered as
$P_N(k) = F_N(k-1) - F_N(k)$,
$k=1,2,\dots,N-1$.
The probability
$F_N(k)$
does not necessarily converge to zero in the limit 
$k \rightarrow \infty$.
Its asymptotic value
$F(\infty)$ 
is equal to the fraction of pairs of nodes in the network 
which belong to different clusters, 
namely for which
$d_{ij} = \infty$.
In fact, $F(\infty)$ can be estimated independently
by using known properties of the fraction of nodes, $g$,
which belongs to the giant component in the asymptotic limit
\cite{Bollobas2001}.
This fraction satisfies 
$g=1-\exp(-cg)$
and 
$F(\infty)=1-g^2$. 
In a finite network
$F(\infty)$ can be replaced by $F_N(N-1)$
since the longest possible distance is $d=N-1$.

In the RSA, one
picks a random node, $i$, as a reference node and examines 
the shell structure of the rest of the network 
around it. 
The number of nodes which are at a distance
$d>k$, $k=0,1,2,\dots,N-1$, from the reference node
is denoted by
$\overline{N}_k$.
The number of nodes at distance $d=k$ 
from the reference node
is denoted by
$N_k$, where $N_0=1$ and 
$N_k = \overline N_{k-1} - \overline N_k$
for $k \ge 1$.
The $N_k$'s obey the recursion equation
$N_{k+1}=\overline N_k(1-q^{N_k})$,
which can be re-written as a second 
order difference equation of the form
$\overline N_{k+1} = \overline N_k q^{\overline N_{k-1}-\overline N_k}$,
where
$\overline N_0=N-1$
and 
$\overline N_1=(N-1)q$.
Using the relation
$\overline N_k = (N-1) \cdot F_N(k)$,
it can be expressed as
\begin{equation}
F_N(k+1) =  F_N(k) q^{(N-1)[F_N(k-1) - F_N(k)]},
\label{eq:daniel}
\end{equation}
\noindent
where
$F_N(0)=1$
and
$F_N(1)=q$.

In the RPA one first picks two distinct
random nodes, $i$ and $j$.
The probability that the distance between them
is larger than $k$ can be related to the probability that
it is larger than $k-1$ by
$F_N(k) = F_N(k-1)P_N(d>k|d>k-1)$,
where
$P_N(d>k|d>k-1)$
is the conditional probability that the distance
is larger than $k$, given that it is larger than $k-1$.
The iteration of this relation yields
\begin{equation}
F_N(k) = F_N(1) \prod_{m=2}^{k} P_N(d>m|d>m-1).
\label{eq:prod}
\end{equation}
\noindent
This means that in order to obtain the distribution $F_N(k)$, all
we need to calculate are the conditional probabilities
$P_N(d>m|d>m-1)$,
for all values of $2 \le m \le k$.

Consider a path of length $k$ 
starting at node $i$ and ending at node $j$
(assuming that there is no such path of length $k-1$ or less).
The path can be decomposed into a single edge from node $i$ to 
an intermediate node $\ell$ and a shorter path of length $k-1$
from $\ell$ to $j$.
Such a path can be ruled out in two ways: either there is no
edge between $i$ and $\ell$ (with probability $q$), 
or, in case that there is such an edge - there is no path of length
$k-1$ between $\ell$ and $j$.
The probability of the latter is 
$P_{N-1}(d>k-1|d>k-2)$,
since the remaining path is embedded in a smaller 
network of $N-1$ nodes.
Combining the two possibilities yields the
recursion equation
\begin{equation}
P_N(d>k|d>k-1)= \left[q + p \cdot P_{N-1}(d>k-1|d>k-2) \right]^{N-2},
\label{eq:recursion}
\end{equation}
\noindent
where the right hand side is raised to the power $N-2$ in order to 
account for all possible ways to choose the intermediate node $\ell$.
In Fig. 1 
we present 
the possible paths of length $k$ between $i$ and $j$. 
This approach follows the spirit of the renormalization group theory
\cite{Binney1993},
since the removal of a node from the network reduces the 
size of the configuration space by a factor of $2^{N-1}$.
This process is repeated $k-1$ times, reducing the network
down to size $N^{\prime}=N-k+1$ and 
closing the recursion equations with
$P_{N^{\prime}}(d>1|d>0) = F_{N^{\prime}}(1)=q$.

\begin{figure}[H]
\begin{center}
\onefigure[width=10cm]{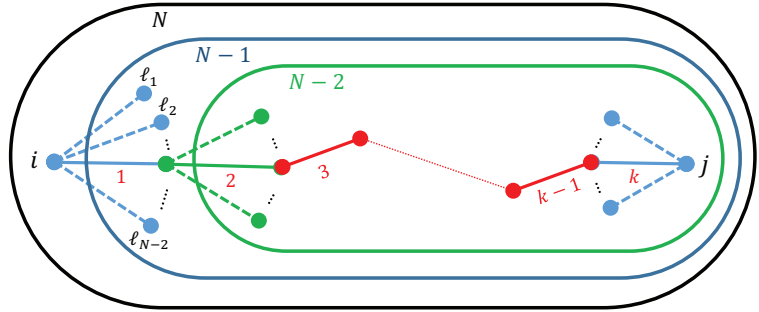}
\caption{
(Color online)
Illustration of the possible paths of 
length $k$ between two random nodes $i$
and $j$ in an ER network of $N$ nodes. 
The first edge of such path connects node $i$ 
to some other node $\ell$, which may be any one 
of the remaining $N-2$ nodes. 
The rest of the path, from $\ell$ to 
$j$ is of length $k-1$ and it 
resides on a smaller network
of $N-1$ nodes.
The path 
of length $k$
from $i$ to $j$ exists only 
when both the edge from $i$ to $\ell$ and the 
path of length $k-1$
from $\ell$ to $j$ exist.
}
\end{center}
\label{fig:1}
\end{figure}

Interestingly,  
inserting $k=2$ in Eq.
(\ref{eq:recursion})
gives rise to the simple and exact expression
\begin{equation} 
P_N(d>2|d>1) = (1-p^2)^{N-2}.
\end{equation}
\noindent
Each path of length $k=2$  between nodes $i$ and $j$
consists of a single intermediate node and two edges.
These paths do not overlap 
and are thus independent.
Paths of lengths $k>2$ may share edges with other paths of the
same length as well as with shorter paths.
Therefore, in the calculation of the 
DSPL
we use conditional probabilities to ensure that no shorter paths exist.
This approach eliminates the correlations between 
paths of different lengths. 
On the other hand, nodes $i$ and $j$ 
may be connected by several paths of the same
length, which may share some edges and thus become correlated.
The RPA does not account for such correlations,
because it assumes that the sub-networks of size $N-1$ are independent.
Averaging over the quenched randomness in each
instance of such network, the RPA provides the distribution 
over an ensemble of networks.

In the limit $p \rightarrow 0$ one can simplify the recursion 
equations and obtain the approximate closed form expression
\begin{equation}
P_N(d>k|d>k-1) = (1-p^k)^{(N-2) \dots (N-k)},
\label{eq:smallp}
\end{equation}
\noindent
for any value of $k$.
This expression is obtained using induction, based on 
Eq. (\ref{eq:recursion})
and the exact
result given above for $k=2$.
This can be understood intuitively since the total
number of possible paths of length $k$ 
between nodes $i$ and $j$ is given by the product
${(N-2) \dots (N-k)}$,
and the probability for each of these paths to be
connected is given by
$p^k$.
This approximation breaks down for values of $p$
which are not exceedingly small, where the correlations
between different paths build up and cannot be ignored.

The regime of sparse networks was studied extensively, 
focusing on the diameter 
(namely, the largest distance between any pair of nodes)
of the giant cluster,
which scales like a constant times $\ln N$,
where the constant is $1/\ln c - 2/\ln c'$, where $c'<1$ satisfies the equation $c' \exp(-c')=c \exp(-c)$ \cite{Riordan2010}.
In the strongly connected regime,
we focus on the case in which 
$p = b N^{\alpha-1}$,
where $b>0$ and $0 < \alpha < 1$.
In this case the average degree increases
with the network size as
$N^{\alpha}$.
We will now derive an asymptotic result for the 
limit
$N \rightarrow \infty$.
In this limit 
$p \rightarrow 0$ 
and therefore the simplified results 
of Eq. (\ref{eq:smallp})
can be used.
Plugging the scaling of $p$ vs. $N$
into  
Eq. (\ref{eq:smallp})
one obtains
\begin{equation}
P_N(d>k|d>k-1) \simeq \left(1 - \frac{b^k}{N^{k(1-\alpha)}}\right)^{N^{k-1}}.
\end{equation}
\noindent
For $N \rightarrow \infty$,
$P_N(d>k|d>k-1) \rightarrow P(d>k|d>k-1)$, where
$P(d>k|d>k-1) = 1$ for $k < 1/\alpha$,
$\exp{(-b^{1/\alpha})}$ for $k = 1/\alpha$
and $0$ for $k > 1/\alpha$.
\noindent
Note that the second case in the above equation 
is obtained only in the special case of
$\alpha = 1/r$, 
where $r$ is an integer. 
Therefore, we will first consider the generic case in which
$\alpha$ is not an exact inverse of an integer.
Inserting the result for the conditional probabilities 
into 
Eq. (\ref{eq:prod})
we obtain
\begin{equation}
P(k) = \left\{
  \begin{array}{lr}
    1 &  : k = \lfloor \frac{1}{\alpha}  \rfloor + 1 \\
    0 &  :  {\rm otherwise},
  \end{array}
\right.
\label{eq:real}
\end{equation}
\noindent
where $\lfloor x \rfloor$ is the integer part of $x$.
In case that 
$\alpha=1/r$ 
we obtain that
\begin{equation}
P(k) = \left\{
  \begin{array}{lr}
    1 - e^{-b^r} & : k = r \\
   e^{-b^r} & : k = r+1 \\
    0 & :  {\rm otherwise}.
  \end{array}
\right.
\label{eq:integer}
\end{equation}
\noindent
These results can be understood intuitively using 
the following argument.
Starting from node $i$, 
we define the shell 
of radius $d=1$ around it
as the set of nodes which are directly connected to $i$. 
The expected value for the
number of nodes in this shell is 
$N_1 \sim N^{\alpha}$.
Proceeding by induction, 
the shell of radius $d$ 
is denoted as the set of nodes
which are directly connected to nodes in the shell
of radius $d-1$.
Thus, the number of nodes in the shell of 
radius $d$ is given by
$N_d \sim N^{d \alpha}$. 
In the asymptotic limit, 
as long as $d \alpha < 1$, the 
shell of radius $d$ still 
consists of an exceedingly small  
fraction of the nodes in the network. 
On the other hand, once 
$d \alpha > 1$, 
this shell includes
almost every node in the network. 
This means that almost all the nodes in the
network are at a distance 
$d = \lfloor 1/\alpha \rfloor +1$ 
from node $i$. 
Since node $i$ was chosen at random, 
this means that the shortest path between almost any 
pair of nodes in the network 
is of length $d$. 

The case of 
$\alpha = 1/r$, 
where $r$ is an integer,
requires a special consideration. 
Based on the argument presented
above, the neighborhood
of radius $d=r$ from node $i$ should include 
all the $N$ nodes. 
However, this counting includes duplications, namely 
nodes which are connected to node $i$ by
several paths of length $r$. 
As a result, there are other nodes which are 
not reached by any of these paths. 
Since the number of nodes of distance 
$r$ from node $i$ scales with $N$, it is clear
that each one of the remaining nodes is connected to at 
least one of them. 
Therefore, the remaining
nodes are at a distance $d=r+1$ from node $i$.

Before presenting the results obtained from the two approaches, 
we refer to an earlier study of
the DSPL
in ER networks 
\cite{Blondel2007}. 
We briefly summarize their approach, 
adapting the notation where appropriate.
The expectation value 
for the number of nodes at a distance $k-1$ or less from the 
reference node is given by:
$n(k) = [1 - F_N(k-1)]N$.
This is due to the fact that the probability for a random node 
to be at a distance smaller than $k$ is $(1-F_N(k-1))$, and 
multiplying by $N$ one obtains $n(k)$. 
In order for a node to be at a distance larger than $k$ from 
the reference node, it must not be directly connected to any 
of the $n(k)$ nodes which are at distance $k-1$ or less 
from the reference node. Picking a random node, the 
probability that it will 
not be connected to any of these nodes is given by
\cite{Blondel2007}
\begin{equation}
F_N(k) = q^{ [1-F_N(k-1)]N }.
\label{eq:blondel1}
\end{equation}
\noindent
This recursion equation can be iterated, 
starting from $F_N(0)=(N-1)/N$,
to obtain $F_N(k)$ for $k=1,2,...$.
A potential problem with this approach is 
that in the estimation of the probability, $F_N(k)$, 
that a random node will be at distance larger than $k$ 
from the reference node, 
Eq. (\ref{eq:blondel1})
ignores the possibility that 
the random node is already connected to the reference node
by a path of length $k-1$ or less.
This is expected to bias the distribution towards larger distances.

In Fig. 2 we present the tail distribution 
$F_N(k)$ vs. $k$, 
for an ER network 
of $N=1000$ nodes
and $p=c/N$, 
where $c=2.5$,
obtained from numerical simulations
for all pairs of nodes ($\times$)
and for pairs of nodes on the same cluster ($+$).
We also present the theoretical results obtained from 
the RSA
($\square$).
and from the RPA
($\circ$).
The results of 
the RSA
agree well with the numerical results for all pairs, 
except for the limit of large distances where the plateau in $F_N(k)$
is lower than the empirical curve.
It means that this approach underestimates
the fraction of pairs for which 
$d_{ij}=\infty$, which is equal to $F(\infty)$.
The results of the RPA
agree well with the numerical results for pairs 
which reside on the same cluster.
This is due to the fact that this approach
reconstructs the remaining network at each iteration
of the recursion equations.
As a result, the quenched randomness of the connectivities
in each realization of the network is annealed,
eliminating the isolated nodes and the small, isolated clusters.
In the RSA there is no such annealing.
Therefore, the RSA applies to all
pairs of nodes in the network while the RPA
applies to pairs of nodes on the same cluster. In the limit of dense networks
there are no isolated components and the two approaches coincide.  

\begin{figure}[H]
\begin{center}
\onefigure[width=10cm]{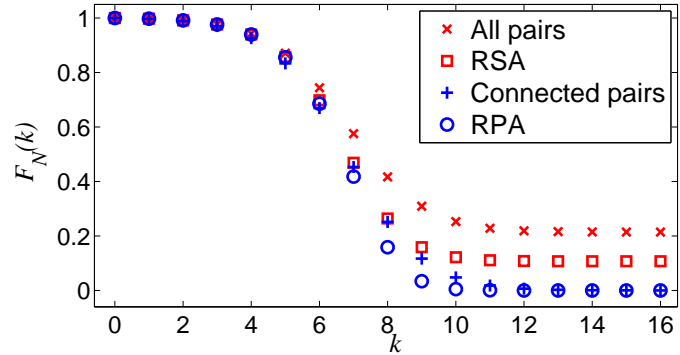}
\vspace{-0.7in}
\caption{
(Color online)
(a) The tail distribution 
$F_N(k)$, vs. $k$ 
for the ER($N,c/N$) network with $N=1000$ and $c=2.5$,
obtained from numerical simulations for all pairs of nodes
($\times$)
and for pairs of nodes on the same cluster ($+$).
The results of the RSA
($\square$)
agree well with the numerical results for all pairs of nodes,
except for the asymptotic tail.
The results of the RPA
($\circ$)
agree well with the numerical results for pairs of nodes on the same 
cluster.
}
\end{center}
\label{fig:2}
\end{figure}

The distribution 
$P_N(k)$
can be characterized 
by its moments. 
The $n$th moment, $\langle k^n \rangle$,
can be obtained 
using the tail-sum formula
$\langle k^n \rangle = \sum_{k=0}^{N-1} [(k+1)^n - k^n] F_N(k)$.
In particular, the first moment is given by
$\langle k \rangle = \sum_{k=0}^{N-1} F_N(k)$
and the second moment by 
$\langle k^2 \rangle = \sum_{k=0}^{N-1} (2k+1) F_N(k)$.
The width of the distribution can be characterized by the
variance 
$\sigma^2 = \langle k^2 \rangle - \langle k \rangle^2$.
Related topological indices 
\cite{Dehmer2014}
such as the Wiener index 
\cite{Wiener1947}
and the Harary index
\cite{Mihalic1992,Ivanciuc1993,Plavsic1993}
were studied in the context of chemical graphs.
It was shown that important properties of molecules can be
obtained using such indices for the graphs representing
their structure
\cite{Wiener1947}.

\begin{figure}[H]
\centerline{\includegraphics[width=6.5cm]{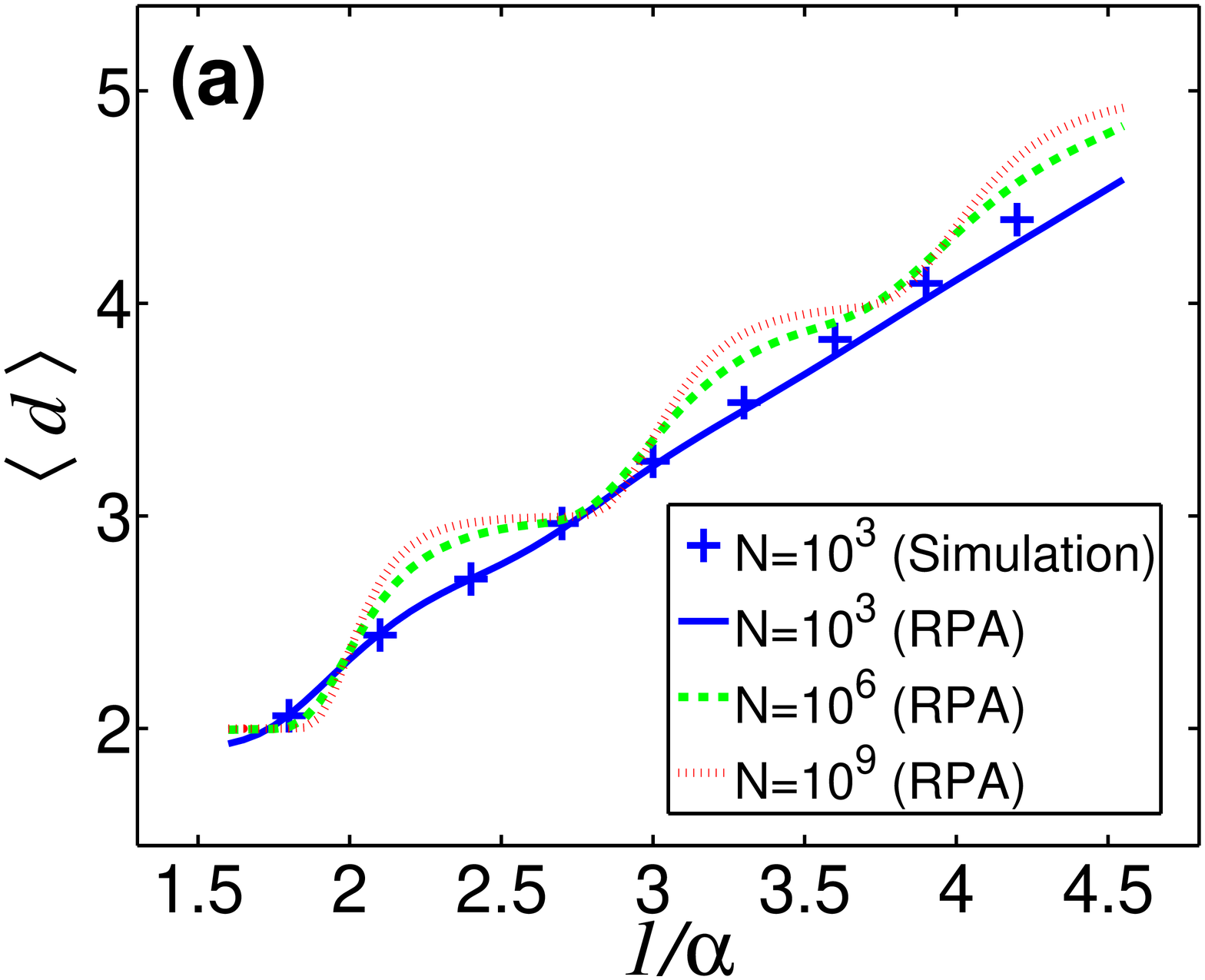},\hspace{0.4in},\includegraphics[width=6.5cm]{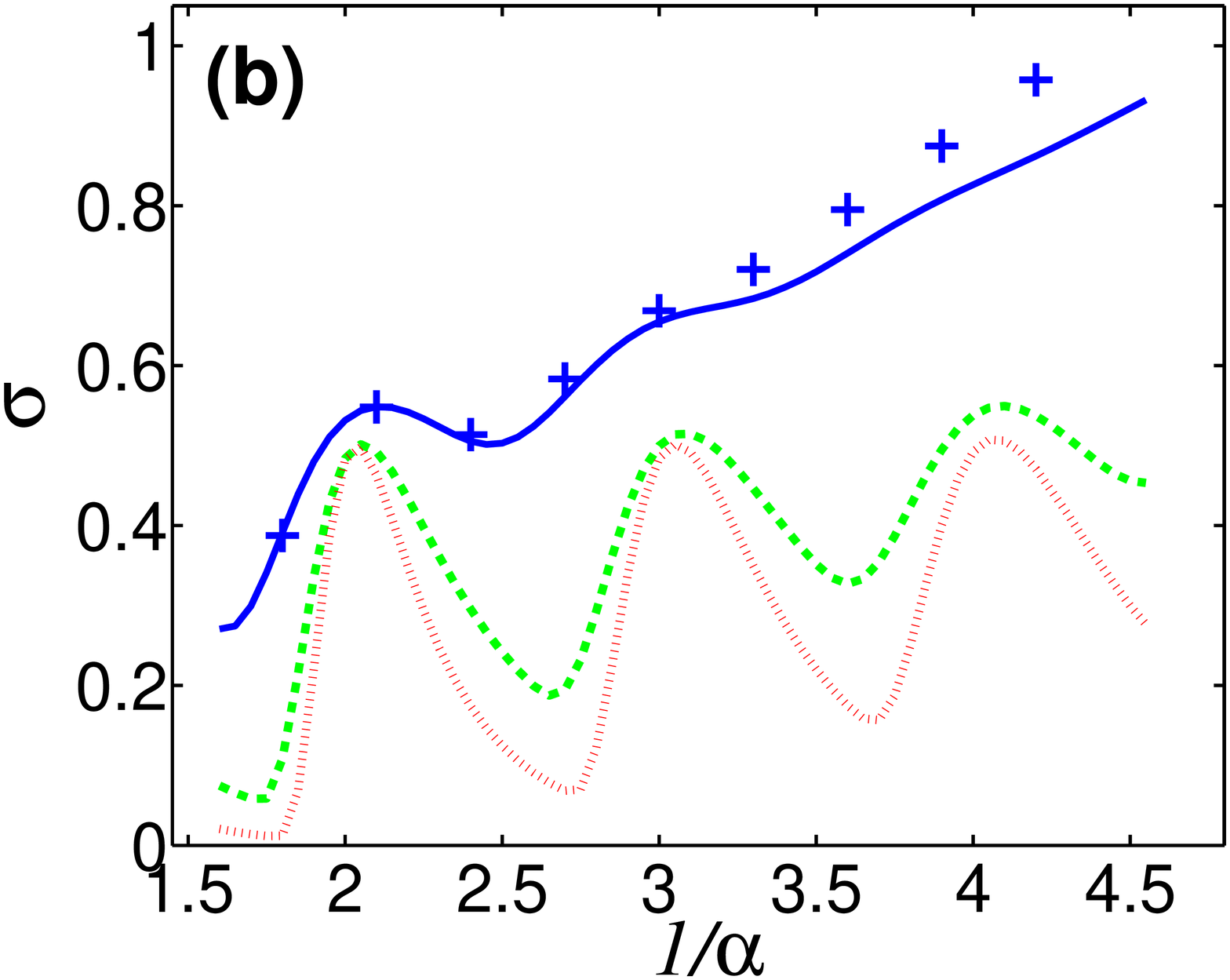}}
\caption{
(Color online)
The average $\langle d \rangle$ (a) 
and the standard deviation $\sigma$ (b)
of the DSPL
in the ER($N,bN^{\alpha-1}$) network vs.
$1/\alpha$ for $b=1$ and 
$N=10^3$ 
(solid line)
$10^6$ 
(dashed line)
and 
$10^9$
(dotted line),
obtained from the RPA.
It is observed that 
$\langle d \rangle \simeq \lfloor 1/\alpha \rfloor +1$,
decorated by a rounded step function,
while $\sigma$ exhibits oscillations with maxima at
integer values of $1/\alpha$.
}
\label{fig:3}
\end{figure}

In Fig. 3(a) we present the average distance $\langle d \rangle$
between pairs of nodes
vs. $1/\alpha$ in dense ER networks.
Following Eqs. 
(\ref{eq:real})-(\ref{eq:integer}),
these functions converge to a staircase form
as $N \rightarrow \infty$.
In Fig. 3(b) we present the standard deviation $\sigma$
vs. $1/\alpha$. 
For finite networks it exhibits 
oscillations of unit period. 
In the asymptotic limit the peaks become vanishingly narrow 
around the integers.

So far we have studied the DSPL
between
all pairs of nodes in the network. 
Below, we consider a reference node $i$ of a known degree, $m$,
and study the DSPL
between
this node and the rest of the network.
We denote the DSPL
between a random node $i$ of degree $m$ and other random nodes, $j$, by
$F_{N|m}(k)=F_N(k|deg(i)=m)$ 
and the corresponding conditional probability by
$P_{N|m}(d>k|d>k-1)$.
In this case, the first iteration of the recursion equation
takes the form
\begin{equation}
P_{N|m}(d>k|d>k-1) = [P_{N-1}(d>k-1|d>k-2)]^{m},
\label{eq:cond_m}
\end{equation}
\noindent
where the expression on the right hand side is obtained
from Eq. (\ref{eq:recursion}).
In Fig. 4 we present the tail distribution 
$F_{N|m}(k)$ vs. $k$, obtained from numerical simulations
for $m=1$ (+), $3$ ($\times$) and $7$ ($\ast$),
in a dilute ER network of $N=1000$ and $c=2.5$.
Each data point is averaged over $20$ independent realizations of the network.
The results of the RPA for
$m=1$ ($\diamond$),
$3$ ($\square$)
and
$7$ ($\circ$)
are in good agreement with the numerical results.
Clearly, the distribution is  strongly affected by the local 
connectivity of the reference node.
The knee of the distribution 
$F_{N|m}(k)$
(which coincides with the peak of the corresponding 
probability density function)
moves to the left as $m$ is increased.
This means that nodes which are strongly connected
at the local level are closer to the rest of the network
than weakly connected nodes.

\begin{figure}[H]
\begin{center}
\onefigure[width=12cm]{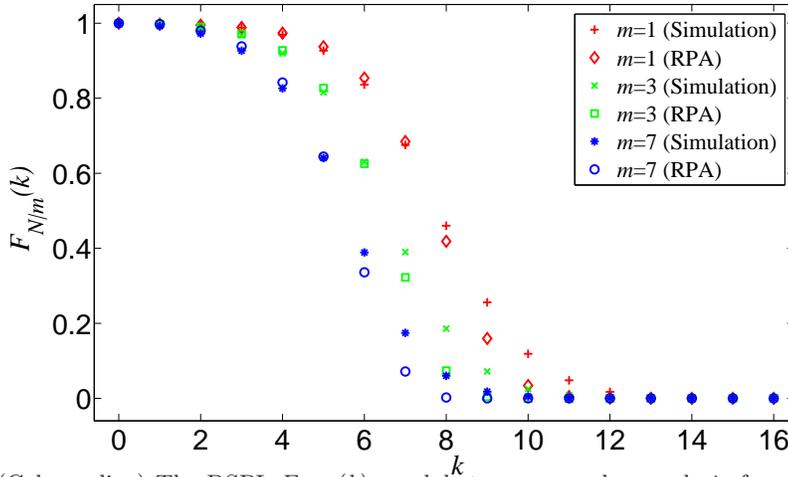}
\vspace{-0.7in}
\caption{
(Color online)
The DSPL 
$F_{N|m}(k)$ vs. $k$ 
between a random node $i$ of a given degree, $m$, and all other nodes 
which reside on the same cluster
in a dilute ER network of $N=1000$ and $c=2.5$.
The results of the RPA for
$m=1$ ($\diamond$),
$3$ ($\square$)
and
$7$ ($\circ$)
are in good agreement with the corresponding numerical results:
$m=1$ (+), $3$ ($\times$) and $7$ ($\ast$).
}
\end{center}
\label{fig:4}
\end{figure}

In summary, we have studied the distribution of shortest path lengths
in ER networks using two complementary theoretical approaches and showed
that they are in good agreement with numerical results.
For large and dense networks the distribution becomes 
extremely narrow and is exactly captured by both approaches.
A slight modification enables us to calculate the DSPL
around a node with a given degree, $m$.
The results exemplify the impact of local features
(such as the degree of a node) on global properties 
(such as the distance distribution) in complex networks.
The proposed theoretical approaches are highly flexible 
and can be applied to more general networks
\cite{Caldarelli2002,Boguna2003}.

\acknowledgments
DbA, OB and PLK thank the Galileo Galilei Institute for Theoretical Physics
in Florence and INFN for the hospitality and support during the workshop on 
Advances in Nonequilibrium Statistical Mechanics in Spring 2014,
where preliminary work was performed.
MN is grateful to the Azrieli Foundation for the award of 
an Azrieli Fellowship.

\end{document}